\documentclass[aps,prl,twocolumn,superscriptaddress]{revtex4-1}

\usepackage{graphicx}
\usepackage{epstopdf}
\usepackage{amsmath}
\usepackage[usenames,dvipsnames]{color}
\usepackage{adjustbox}
\usepackage{tabularx}

\usepackage[normalem]{ulem}

\def\d{{\rm d}}

\begin{document}

\title{Origin of the Bauschinger Effect in Amorphous Solids}

\author{Sylvain Patinet}
\author{Armand Barbot}
\author{Matthias Lerbinger}
\author{Damien Vandembroucq}
\affiliation{PMMH, CNRS, ESPCI Paris,  Universit\'e PSL, Sorbonne Universit\'e,
 Universit\'e de Paris, 75005 Paris, France}
\author{Ana\"{e}l Lema\^{i}tre}
\email[]{anael.lemaitre@enpc.fr}
\affiliation{Navier, Ecole des Ponts, Univ Gustave Eiffel, CNRS, F-77420 Marne-la-Vall\'ee, France}

\date{\today}

\begin{abstract}
We study the structural origin of the Bauschinger effect by accessing numerically the local plastic thresholds in the steady state flow of a two-dimensional model glass under athermal quasistatic deformation. More specifically, we compute the local residual strength, $\Delta\tau^{c}$, for arbitrary loading orientations and find that plastic deformation generically induces material polarization, i.e., a forward-backward asymmetry in the $\Delta\tau^{c}$ distribution. In steady plastic flow, local packings are on average closer to forward (rather than backward) instabilities, due to the stress-induced bias of barriers. However, presumably due to mechanical noise, a significant fraction of zones lie close to reverse (backward) yielding, as the distribution of $\Delta\tau^{c}$ for reverse shearing extends quasilinearly down to zero local residual strength. By constructing an elementary model of the early plastic response, we then show that unloading causes reverse plasticity of a growing amplitude, i.e., reverse softening, while it shifts away forward-yielding barriers. This result in an inversion of polarization in the low-$\Delta\tau^{c}$ region and, consequently, in the Bauschinger effect. This scenario is quite generic, which explains the pervasiveness of the effect.
\end{abstract}


\maketitle

The Bauschinger effect~\cite{bauschinger_uber_1886} is the remarkably
common property that after experiencing plastic strain, materials generally exhibit a softer stress response under reverse loading as compared with reloading. Initially observed in mono- and polycrystalline metals~\cite{Buckley-ActaMetal56,Asaro-ActaMetal75}, this
phenomenon has been evidenced in
polymers~\cite{SendenDommelenGovaert2010}, and more recently in
amorphous materials such as metallic
glasses~\cite{deng_simulations_1989,*Greer-JAC14}. It
is thus found in almost all material classes. Yet, its origin remains
the topic of ongoing debates across the concerned disciplines.

Our interest here is to understand the origin of the Bauschinger
effect in amorphous solids. This is an especially challenging goal since there is no consensus today on how to describe the
internal state of a glass~\cite{spaepen_microscopic_1977,*shi_evaluation_2007,sollich_rheology_1997}
in view of predicting its mechanical response.
The plastic response of glasses is known to
result from local rearrangements (or
``flips'')~\cite{Argon1979,falk_dynamics_1998,schuh_atomistic_2003,MaloneyLemaitre-PRL04a,*MaloneyLemaitre2006} that occur when certain regions (``zones'') a few atoms wide
reach local instabilities. Yet, due to structural disorder, these instabilities are not associated with specific local structures such as topological defects;
they may also occur at different local yield stress levels~\cite{tsamados_study_2008,patinet_connecting_2016}.
Besides, every zone flip introduces long-range, elastic, stress fluctuations, which act as a
mechanical noise, that may cause secondary events and avalanche
behavior~\cite{TalamaliPetajaVandembroucqRoux2011,MaloneyLemaitre-PRL04a,*MaloneyLemaitre2006}.
Any stable packing, hence, approaches instabilities haphazardly~\cite{LemaitreCaroli2007} as its local stress
fluctuates under the combined effects of external forcing and mechanical noise.


Few numerical works exist on the Bauschinger effect in amorphous
solids~\cite{frahsa_bauschinger_JCP2013,KarmakarLernerProcaccia2010a,rountree_plasticity-induced_2009,rodney_yield_2009}. Procaccia and co-workers found a signature of loading asymmetry in high order derivatives of the potential energy
surface~\cite{KarmakarLernerProcaccia2010a}. In a model of silica, in unloaded states after shear plasticity, Rountree \textit{et al.}~\cite{rountree_plasticity-induced_2009}
  observed the emergence of a type of structural
  anisotropy captured by a variant of the fabric tensor~\cite{Kuo-Geotechnique98,*Radjai-inbook04}
  classically associated with structural asymmetry in granular materials~\cite{Rothenburg-IJSS04,*radjai_modeling_2017}.
  Such findings, however,
remain difficult to relate to a physical picture of flow
mechanisms in the spirit of mesoscale or mean-field
models~\cite{sollich_rheology_1997,falk_dynamics_1998,Hebraud-Lequeux-PRL98,Nicolas-RMP18}. Rodney
and Schuh~\cite{rodney_yield_2009} used the ART
method~\cite{Mousseau-PRE00} to sample the barriers of a sheared
system; they found a signature of polarization in the
strains associated with barrier crossings~\cite{ArgonKuo-MSE79},
but could not connect it directly to the Bauschinger effect.

Here, we identify the origin of the Bauschinger effect in an amorphous solid under steady athermal quasistatic (AQS) flow~\cite{MaloneyLemaitre-PRL04a,MaloneyLemaitre2006}, using a recently developed method, which consists in probing the instabilities of small circular domains under strain~\cite{puosi_probing_2015,patinet_connecting_2016}, and was recently extended to deal with deviatoric strains of arbitrary orientations~\cite{barbot_local_2018}. By measuring $\Delta\tau_c$, the local residual strength, for arbitrary strain orientations, we bring evidence of a strain-induced material polarization, which we characterize precisely both in steady state and during the Bauschinger test.

Our analysis shows that the Bauschinger effect originates from an inverse polarization of the low-$\Delta\tau_c$ tails during unloading. More precisely, we find that, in steady state, the distribution of $\Delta\tau_c$ extends down to $\Delta\tau_c=0$ for any strain orientation, a property expected to arise due to mechanical noise. Strikingly, this holds even for barriers responding to reverse shearing. It follows that, from its very onset, unloading causes reverse plasticity of a growing amplitude, i.e., reverse softening, while shifting forward-yielding barriers away. The Bauschinger ensues since after any finite amount of unloading, the reverse response (i.e. the continuation of unloading) is soft (plastic), while reloading is nearly elastic.

This work uses the same numerical system as
Ref.~\cite{barbot_local_2018,BLLVP_SB}: a two-dimensional (2D) binary Lennard-Jones
model with second order smoothing near the interaction
cutoff. Physical units are fixed by the characteristic energy and
length scales of the pair potential. The simulation cell is square and
periodic, of fixed volume. We use $10^4$-atom configurations and
systematically collate 100 independent runs to obtain statistically
significant data. Plastic deformation is applied in simple shear, with
Lees-Edwards boundary conditions, using the AQS protocol, in which a
system is deformed by small increments of affine strain $\Delta\gamma_{xy}=10^{-4}$ followed by
energy minimization, which guarantees mechanical
balance~\cite{MaloneyLemaitre-PRL04a,*MaloneyLemaitre2006}. As a
result, the system tracks reversible elastic branches except at
instabilities where avalanchelike plastic events occur and dissipate
energy.

As is well known, the early shear response of a glass depends significantly on its preparation, and especially on its degree of relaxation: a poorly relaxed glass typically displays strain hardening; a very well relaxed glass typically develops a peak stress followed by softening and usually accompanied by localization. Yet, when steady shear can be maintained beyond the initial, transient, response, all glasses are eventually driven toward a unique ensemble. This {steady flow state} is usually inaccessible in experiments on hard glasses due to strain localization, but is commonly observed in soft glasses, and can be easily realized in numerical simulations using periodic boundary conditions. This is illustrated in the Supplemental Material~\cite{SM}, where we monitor the convergence of shear stress to a unique level, starting from three widely different glasses, namely prepared by instantaneously quenching a high temperature liquid (HTL), an equilibrated supercooled liquid (ESL), or a system relaxed via a slow gradual quench (GQ)~\cite{barbot_local_2018}.

\begin{figure}[b]
   \includegraphics[width=0.95\columnwidth]{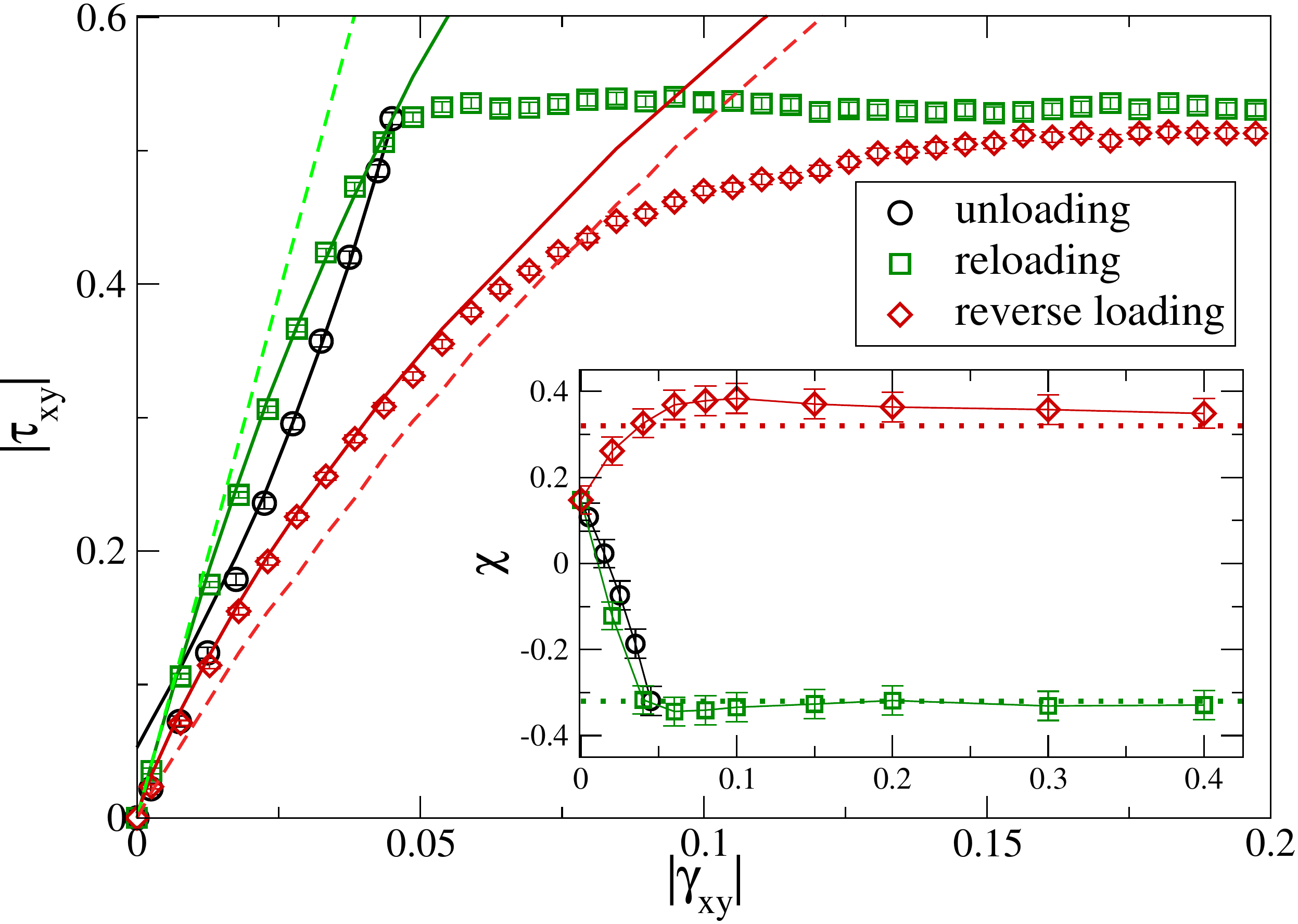}
\caption{\label{fig:bauschinger} Mean stress vs strain (both in
  absolute values) during three tests: unloading from steady flow (black);
  backward (red) and forward (green) loading from fully unloaded (zero stress) configurations. In all three cases, strain is measured with reference to the zero-stress state.
  Solid black line: fit of the unloading curve using Eq.~(\ref{eq:model}) and $\rho\, a^2\Delta\epsilon_0\simeq 0.25$.
  Solid green and red lines: consequent predictions for the Bauschinger tests.
  Dashed lines: predicted reloading and backward loading curves, when the model is used starting from the steady flow barrier distribution (see text), i.e., while taking into account its prediction for the unloading-induced asymmetry. Inset: corresponding evolution of mean barrier polarizations and their asymptotic values (dotted lines).
  }
\end{figure}

Here, as detailed in Fig.~\ref{fig:bauschinger}, we evidence the Bauschinger effect starting from steady flow configurations, so as to emphasize that it is unrelated to the preparation-dependent, hardening or softening, transient response of the glass. In this figure, the origin of strains is taken with reference to the zero-stress states reached after unloading steady flow configurations.
The unloading stress-strain relation (black) appear nearly, but not quite, elastic. A
small (less that $\simeq0.6\%$) but clear hysteresis is seen when
reloading (green), which entails that unloading induces
a small amount of plasticity. The
reverse loading curve (red), which is the continuation of unloading beyond zero stress, i.e., at negative strains, is considerably softer than the
reloading (green) one---this is the Bauschinger effect.

To understand the origin of this phenomenon, we now analyze
steady flow configurations using the method of
Refs.~\cite{patinet_connecting_2016,barbot_local_2018}. It consists in
identifying the first plastic event undergone by atoms inside a small
circular test domain (of radius $R_{\rm free}=5$) when forced by imposing
an affine strain to the outer
atoms within a shell of width larger than the pair interaction cutoff
$R_{\rm cut}$. We only consider pure (deviatoric) strains parametrized as
\begin{equation}
  \propto\begin{pmatrix}-\sin2\alpha & \cos2\alpha\\ \cos2\alpha & \sin2\alpha\end{pmatrix}
  \end{equation}
with a positive prefactor and $2\alpha\in[0,2\pi]$. Thus
$2\alpha=0$ when strain is aligned with the simple shear flow
direction, and $2\alpha=\pi$ for reverse
loading.
Statistically significant data are accumulated by considering
all inclusions centered on regular grid points with a mesh size
$\approx R_{\rm cut}$, while $2\alpha$ takes values at regular
($\pi/9$) intervals.

\begin{figure}[tb]
\begin{center}
\begin{minipage}[b]{0.47\columnwidth}
\includegraphics[width=0.95\columnwidth]{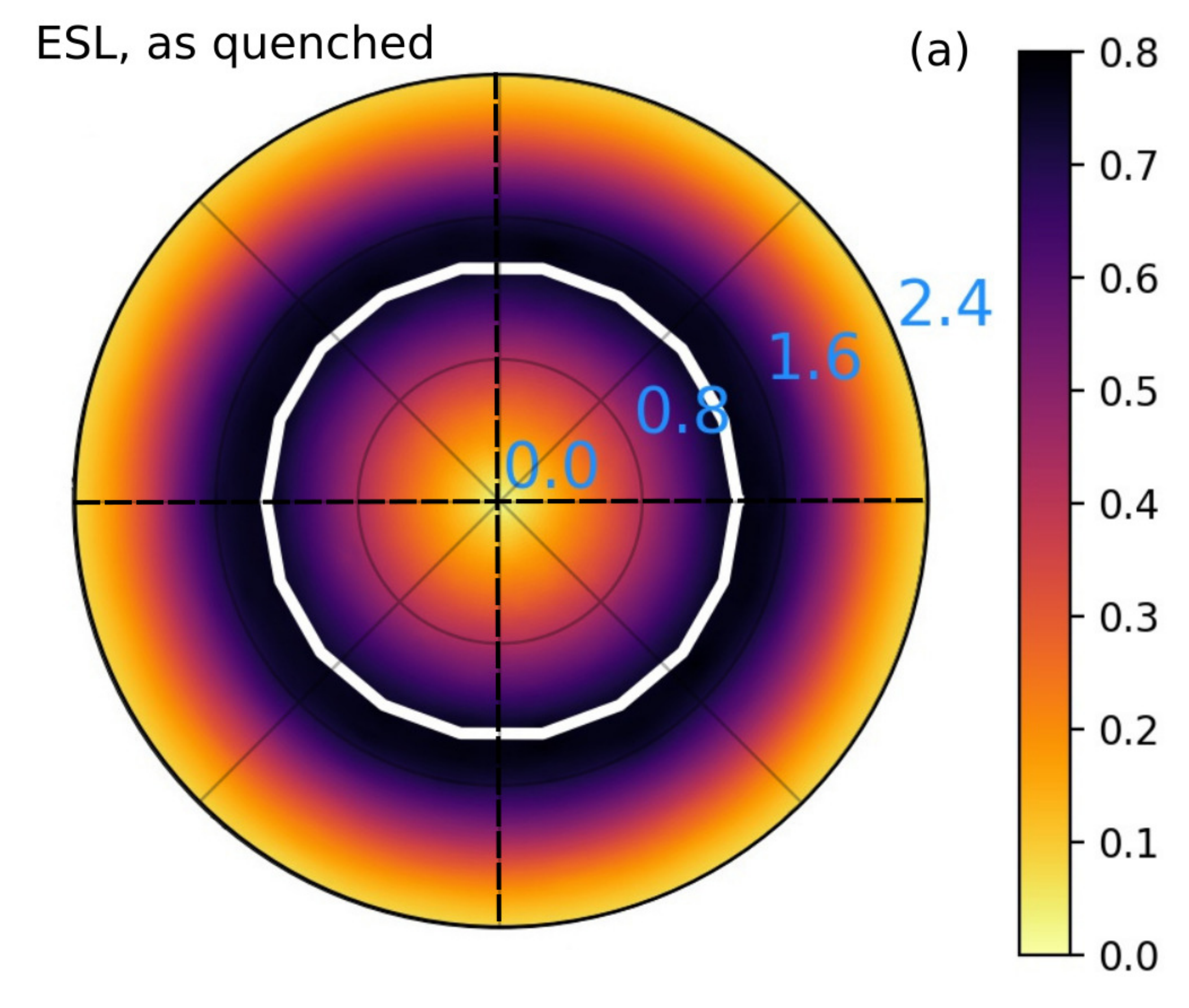}
\includegraphics[width=0.95\columnwidth]{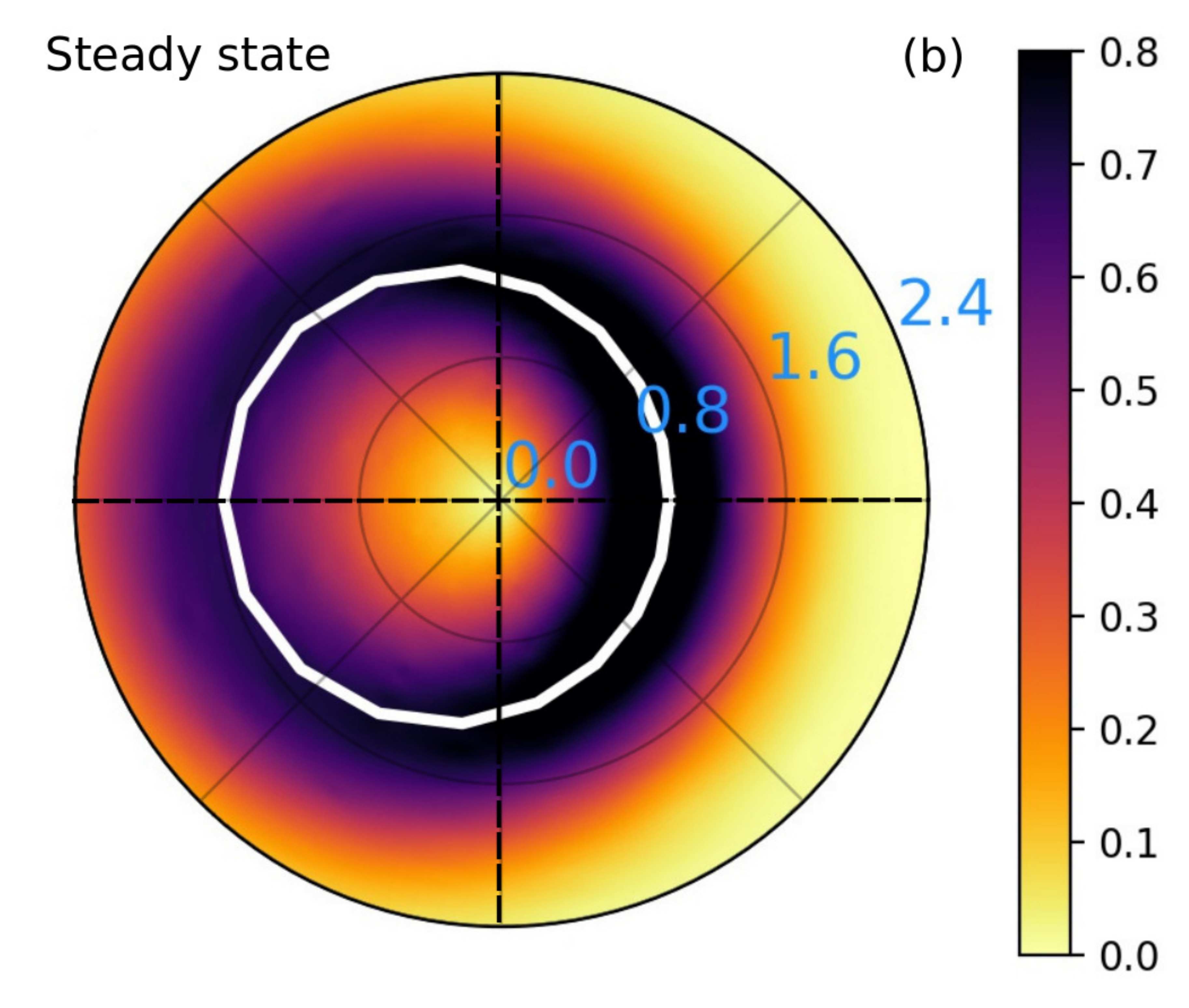}
\includegraphics[width=0.95\columnwidth]{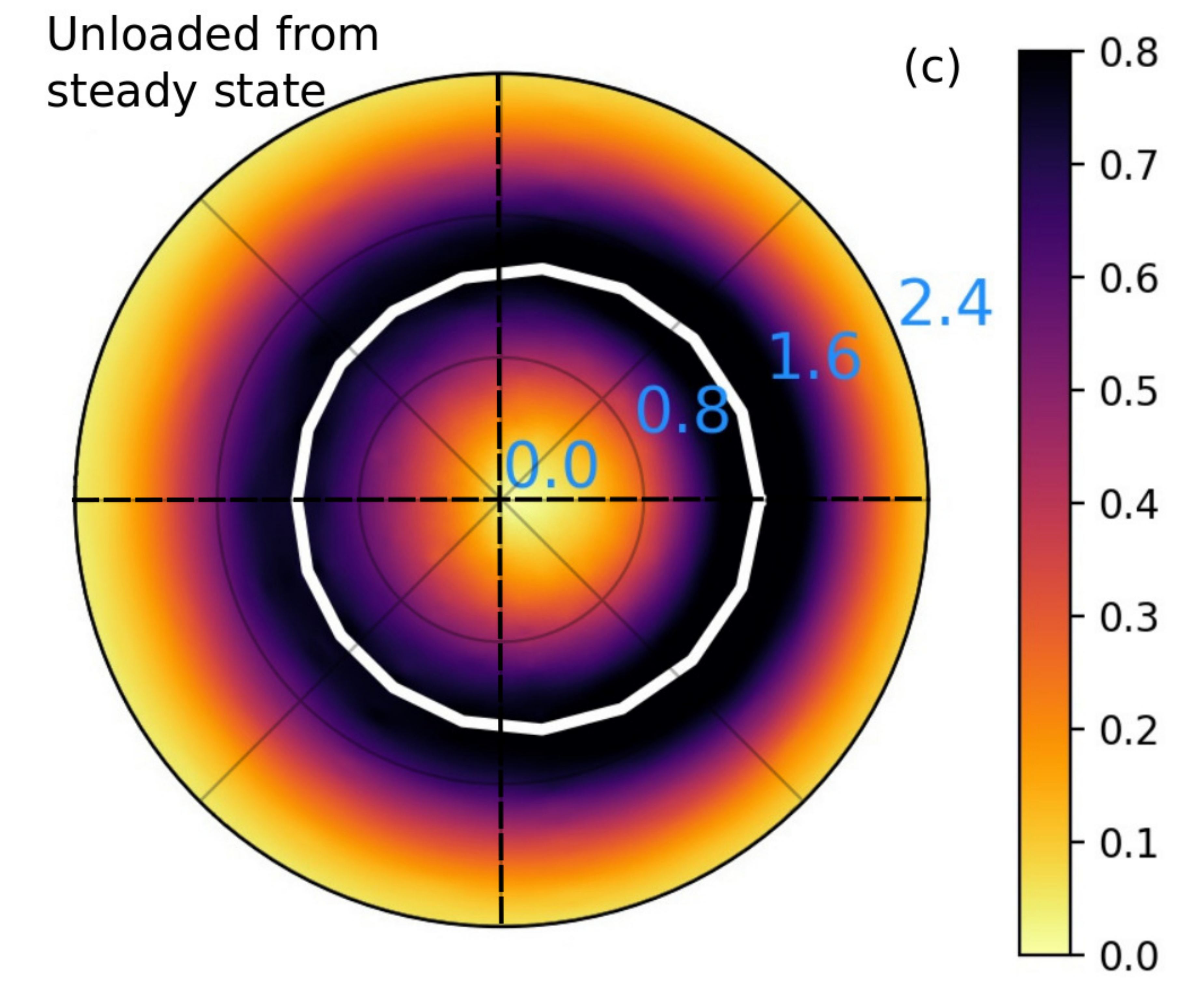}
\end{minipage}
\hfill
\begin{minipage}[b]{0.515\columnwidth}
\includegraphics[width=0.95\columnwidth]{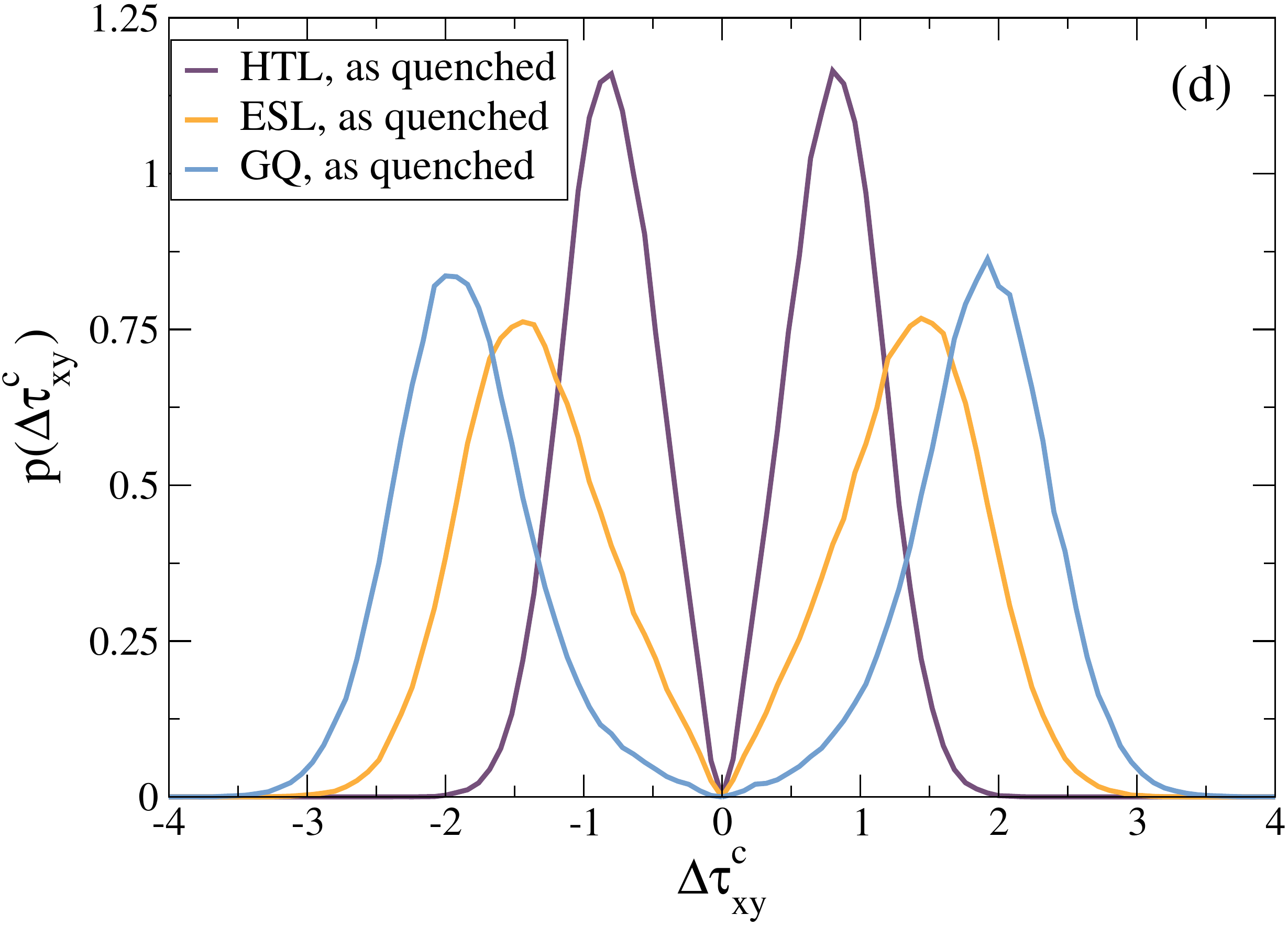}\\[2mm] 
\includegraphics[width=0.95\columnwidth]{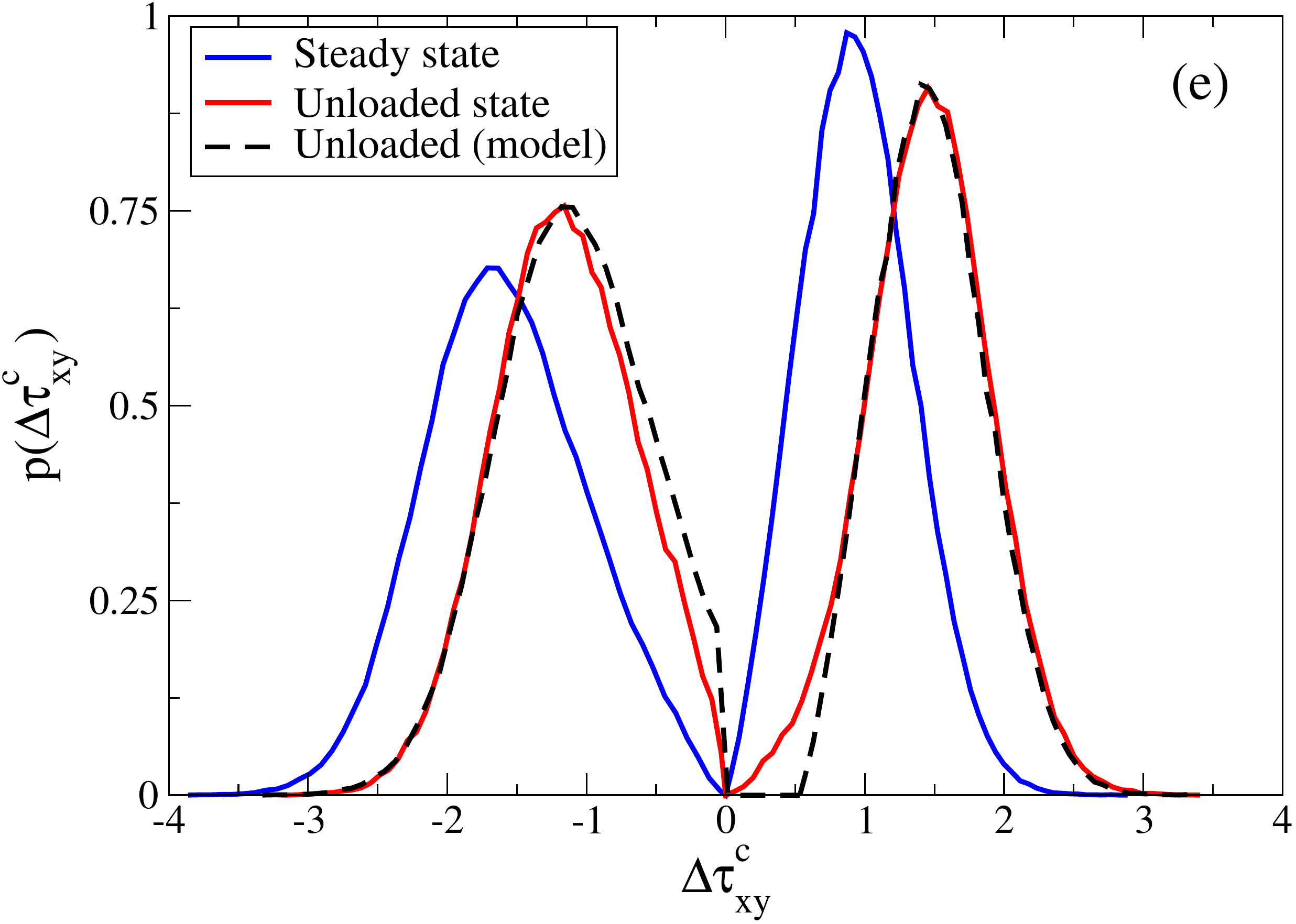}\\[2mm] 
\includegraphics[width=0.95\columnwidth]{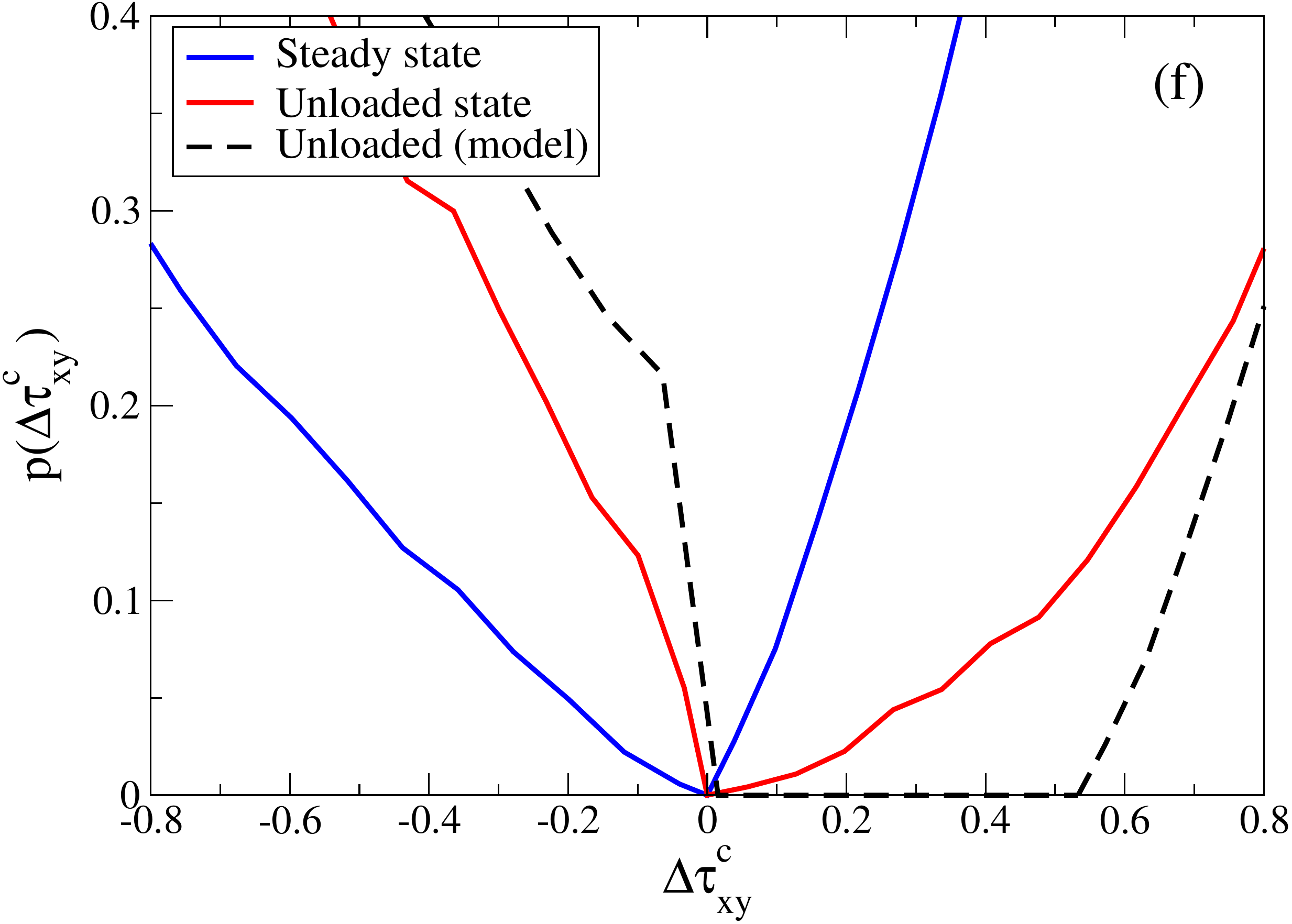} 
\end{minipage}
\end{center}
\caption{\label{fig:barriers} Left frames: the polar function
  $P(\Delta\tau^c,2\alpha)$,
  with $\langle\Delta\tau^c\rangle(2\alpha)$ in white. (a)
  As-quenched isotropic (ESL) state, (b) steady state, and (c) unloaded steady state configurations.
  Right frames: cuts of $P(\Delta\tau^c;2\alpha)$ along the
  forward ($2\alpha=0$) and backward ($2\alpha=\pi$) directions (d) in our three as-quenched glasses of different degrees of relaxation (HTL, ESL and GQ)~\cite{barbot_local_2018}. (e,f) In steady flow (blue) and unloaded states (red),
  and (up to a scaling factor) model prediction for unloading (see text, dashed black).
}
\end{figure}

For each test domain and each $2\alpha\in[0,2\pi]$, we measure the
average (over inclusion atoms) shear stress
conjugate to the imposed pure shear deformation.
The residual strength of the inclusion in this orientation, $\Delta\tau^c(2\alpha)$, is the
corresponding stress increment at the first instability.
The left panels of Fig.~\ref{fig:barriers} display polar maps of the function $P(\Delta\tau^c;2\alpha)$ in the $(\Delta\tau^c,2\alpha), 2\alpha\in[0,2\pi]$ plane.
The right panels show cuts of this function along the $x$ axis, i.e. plots of $P$ vs $\Delta\tau_{xy}^c=\Delta\tau^c$ for $2\alpha=0$ (forward), and vs $\Delta\tau_{xy}^c=-\Delta\tau^c$ for $2\alpha=\pi$ (backward).

In as-quenched systems, as illustrated in Fig.~\ref{fig:barriers}-(a) for the ESL, $P(\Delta\tau^c;2\alpha)$ is isotropic. Moreover [see cuts on panel~(d)], the more relaxed the system, the higher the local yield stresses, as previously observed~\cite{barbot_local_2018}.


In the steady flow ensemble [Fig.~\ref{fig:barriers}(b) and blue curves in panels~(e),(f)], $P(\Delta\tau^c;2\alpha)$ is clearly anisotropic, and more precisely \emph{polarized} in the sense that it breaks the right-left, $\cos2\alpha\to-\cos2\alpha$ symmetry, corresponding to the sign inversion of the off-diagonal strain. The mean barrier height $\langle\Delta\tau^c\rangle(2\alpha)$, in white in Fig.~\ref{fig:barriers}(b), remains circular (a curious feature we cannot explain), but is shifted horizontally by $\chi=\frac{1}{2}\left(\langle\Delta\tau^c\rangle(0)-\langle\Delta\tau^c\rangle(\pi)\right)$, a quantity we call the \emph{mean barrier polarization}. Since $\chi\simeq -0.31 <0$, inclusions are on average closer to the forward ($\alpha=0$) barriers: this is, of course, expected since the steady flow ensemble is under a positive average stress $\overline{\tau}_{xy}^{\rm flow}=0.53$.

In unloaded states [Fig.~\ref{fig:barriers}(c)], $\langle\Delta\tau^c\rangle(2\alpha)$ (white) is still circular, but shifted to the right: unloading inverts the mean barrier polarization as $\chi^{\rm unloaded}\simeq0.14>0$; inclusions are then, on average, closer to reverse barriers. We systematically examined $P(\Delta\tau^c;2\alpha)$ and $\langle\Delta\tau^c\rangle(2\alpha)$ at many strain levels (not shown) and always found $\langle\Delta\tau^c\rangle(2\alpha)$ to be hardly distinguishable from a circle, so that $\chi$ is the center of the mean yield curve. It thus closely resembles the ``backstress'' in its initial meaning as a phenomenological parameter meant to represent a strain-dependent shift of the yield surface~\cite{Lemaitre_book_1990}.

In continuum theories of plasticity, although this was recently challenged~\cite{queyreau_origins_2019}, the backstress is often presumed to reflect an asymmetry of local stress~\cite{rice_structure_1970,Asaro-ActaMetal75,*mughrabi_dislocation_1988}. We thus emphasize that, in our system, the unloaded local stress distribution is nearly perfectly symmetric~\cite{SM}. The barrier distribution asymmetry does not result from stress asymmetry, but from the dynamical equilibrium between
the postflip production of new barriers (rejuvenation) and the preferential elimination of forward-yielding ones.

The evolution of $\chi$ with strain is reported in the inset of Fig.~\ref{fig:bauschinger}, for the three considered tests. Clearly, the barrier distribution develops a history-dependent forward-backward asymmetry, which is inverted during unloading. This raises the question of the possible link between mechanical polarization and the Bauschinger effect. Yet, since $\chi$ overshoots its steady state value (dotted lines, inset of Fig.~\ref{fig:bauschinger}), it does not appear sufficient to \emph{characterize} the material state. If it did, the barrier distribution would recover symmetry when $\chi$ vanishes, which is not the case~\cite{SM}, as it can be guessed from Fig.~\ref{fig:barriers}-(e), since certain asymmetric features (narrowness, peak heights) are not inverted after full unloading.

The question remains to understand how the forward-backward asymmetry of the barrier distribution may play a role in the Bauschinger effect. For this purpose, we recall that, in AQS plasticity, yielding occurs when atomic packings are mechanically brought beyond local yield thresholds~\cite{ArgonBulatovMottSuter1995,MaloneyLemaitre-PRL04b,*MaloneyLemaitre2006,LemaitreCaroli2006}. Therefore, the early plastic response during Bauschinger tests (i.e., forward or backward loading from zero stress) is expected to result primarily from the crossing of the nearest thresholds responding to the external forcing orientation, in the initial, unloaded, state. This points to the forward-backward asymmetry of the \emph{small barrier tails} of $P(\Delta\tau^c;2\alpha)$, which [Fig.~\ref{fig:barriers}(f)], like $\chi$, is visibly inverted during unloading.

\begin{figure}[b]
\includegraphics[height=0.45\columnwidth]{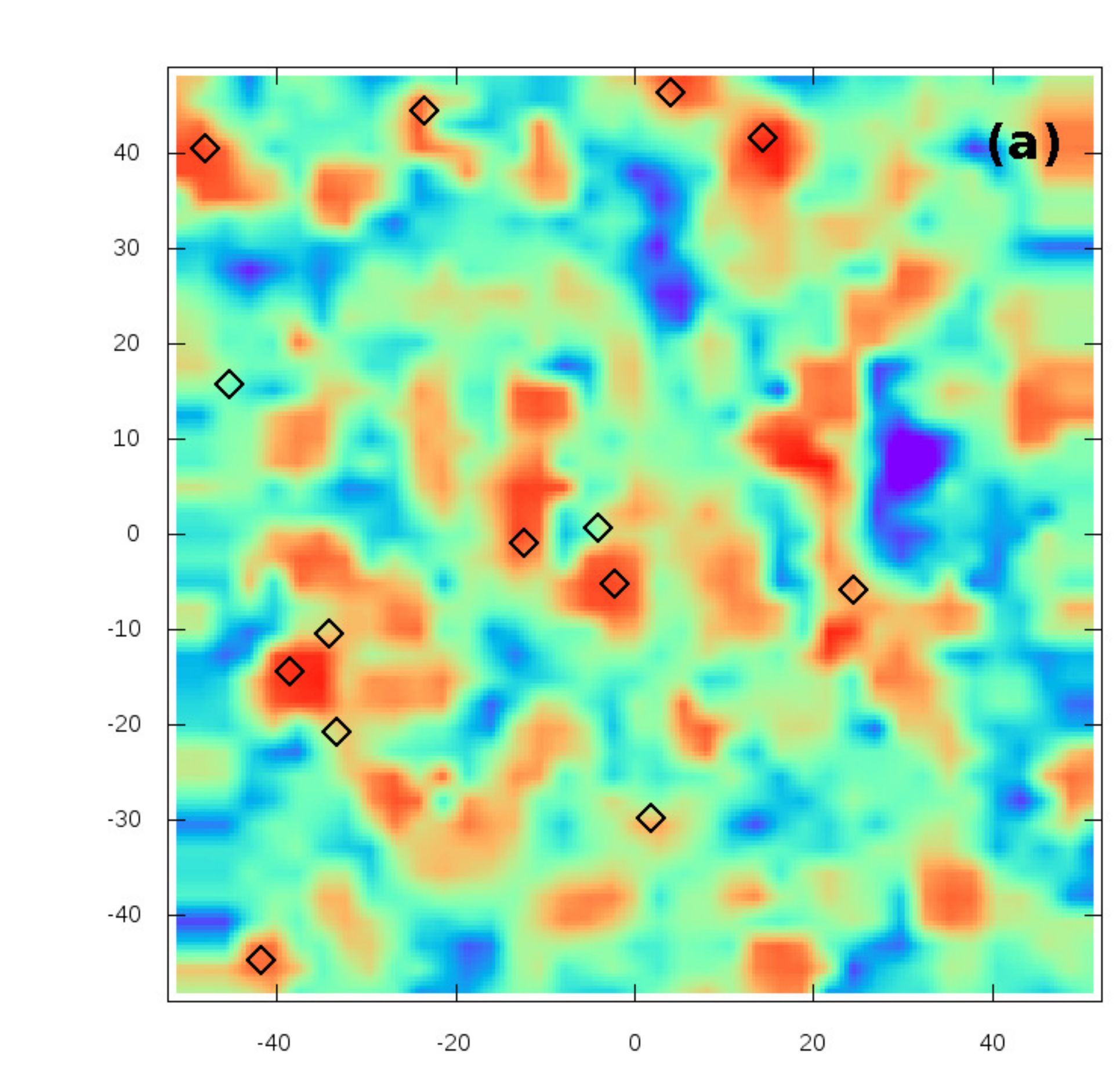}
\includegraphics[height=0.45\columnwidth]{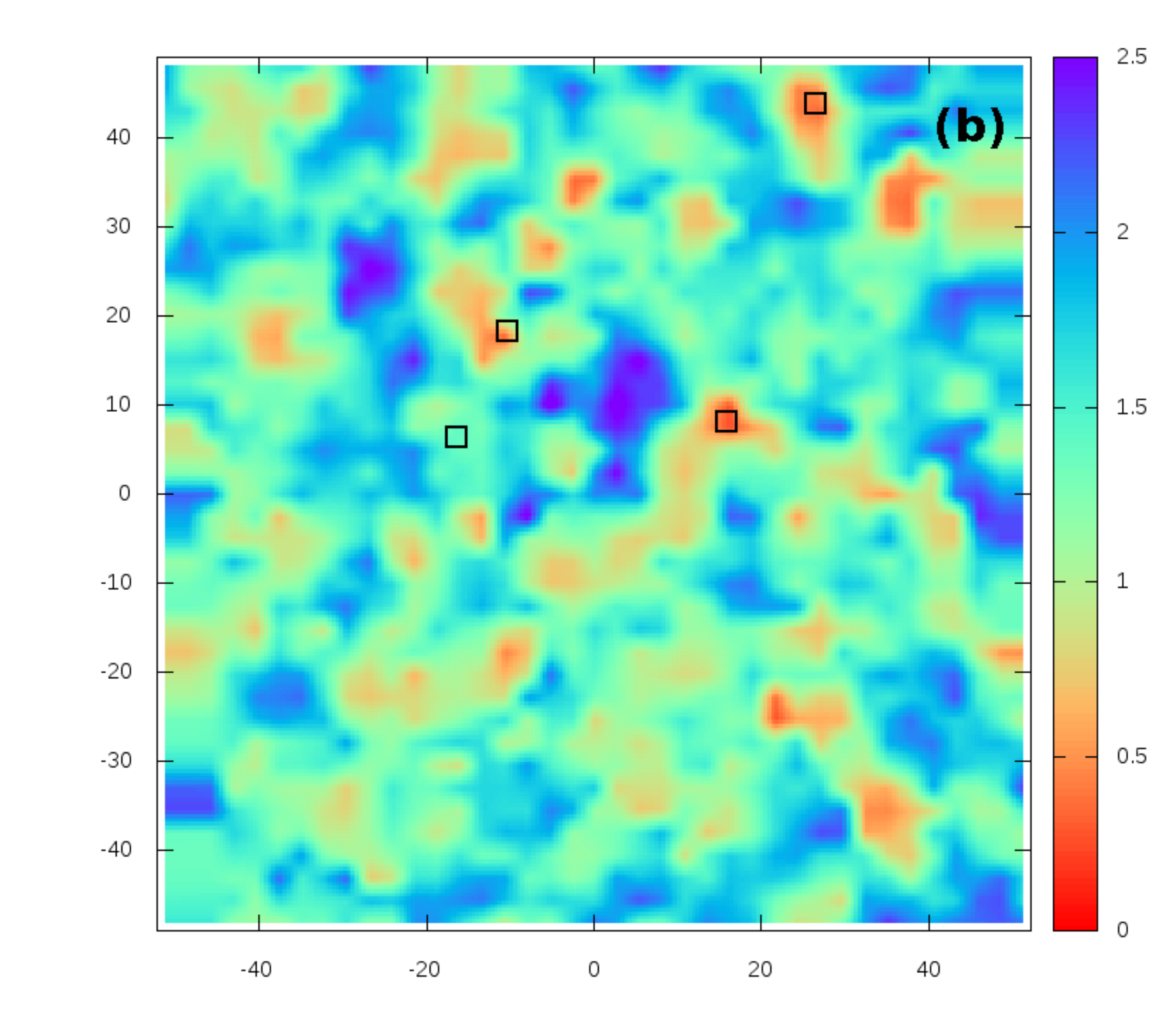}
\caption{\label{fig:maps} Maps of the local residual strength for shearing (a) in the backward ($2\alpha=\pi$) and (b) forward ($2\alpha=0$) directions. Symbols show the loci of plastic events in the 2\% of strain in the corresponding direction.}
\end{figure}

To illustrate the key role of small barriers, we take an arbitrary zero-stress configuration and report in Fig.~\ref{fig:maps} its backward~(a) and forward~(b) barrier maps, on top of which we mark the locations of the local flips undergone in the first 2\% of strain in the corresponding loading direction. Clearly, there is a higher fraction of small barriers (red) and more events (symbols) in panel~(a) rather than~(b). Also, in both cases, the loci of the plastic events seem to correlate with the low barrier regions. This supports our expectation that the forward-backward response contrast (the Bauschinger effect) results from the small barrier density bias in unloaded states.

This idea is quantitatively tested by constructing an elementary model relating the early plastic response to the barrier distribution in the initial state. Three types of loading are considered: both Bauschinger tests, along with unloading from steady state. In all three cases, the macroscopic strain increment $\delta\gamma_{xy}$ is taken with reference to the initial state: it grows positive for reloading, and negative, for unloading and reverse loading. The strain-induced change in macroscopic stress is written as $\delta\overline{\tau}_{xy}=\mu\delta\gamma_{xy}-\delta\overline{\tau}^{\rm pl}_{xy}$ with $\mu\simeq 15.7$ the shear modulus, and $\delta\overline{\tau}^{\rm pl}_{xy}$ the stress released by plastic events.

For all the three considered cases, we looked at plastic drops within the first few percents of strain~\cite{SM}, and found them to typically involve isolated rearrangements, which supports that avalanche dynamics are inactive, and hence mechanical noise can be neglected. Since we seek to capture the beginning of the stress response only, we also neglect rejuvenation. We thus assume that early plasticity results exclusively from instabilities of fixed thresholds $\tau_{xy}^c$, preexisting in the initial state. Neglecting elastic heterogeneities, the preyield local stress reads $\tau_{xy}(\delta\gamma_{xy}) =\tau_{xy}(0)+\delta\overline{\tau}_{xy}$, and yielding occurs when $\Delta\tau_{xy}^c(\delta\gamma_{xy})=\Delta\tau_{xy}^c(0)-\delta\overline{\tau}_{xy}$ vanishes~\footnote{During unloading and reverse loading $\delta\gamma_{xy},\delta\overline{\tau}_{xy}<0$, backward barriers are triggered, and local stability requires $\Delta\tau_{xy}^c<0$; during reloading, $\Delta\gamma_{xy},\delta\overline{\tau}_{xy}>0$, and local stability requires $\Delta\tau_{xy}^c>0$},
so that
\begin{equation}\label{eq:model}
\mu\delta\gamma_{xy}-\delta\overline{\tau}_{xy}=\delta\overline{\tau}^{\rm pl}_{xy}=2\mu\rho a^2\Delta\epsilon_0\,\int_{0}^{\delta\overline{\tau}_{xy}}p\left(\Delta\tau_{xy}^c\right){\rm d} \Delta\tau_{xy}^c
\end{equation}
which defines $\delta\overline{\tau}_{xy}(\delta\gamma_{xy})$.
Here, $p$ is the distribution, in the initial state, of the barriers responding to the considered forcing, $a$ a the typical zone size, $\Delta\epsilon_0$ the typical strain release, and $\rho$ is the number density of yield barriers~\footnote{In our system of volume $V$, the number $N=\rho V$ of barriers responding to local forcing in a given orientation is finite. Note in particular that when our protocol is applied on different, yet overlapping patches, the same barriers are often detected. $\rho$ can be understood a $1/\xi^2$ with $\xi$ the length at which these measurements decorrelate.}.

Equation~(\ref{eq:model}) provides a quantitative test of the relation between barrier tails and the Bauschinger effect because it effectively depends only on the combination $\rho\, a^2\Delta\epsilon_0$, i.e., of a single unknown parameter that can be obtained by fitting (solid black line on Fig.~\ref{fig:bauschinger}) the beginning of the unloading curve (using for $p$ the steady state backward barrier distribution). This yields $\rho\, a^2\Delta\epsilon_0\simeq0.25$. To confirm the relevance of this value, we have computed coarse-grained local strain changes during isolated plastic events, and estimate $a^2\Delta\epsilon_0$ (not shown) to lie in the 0.4--0.7 range. We have also estimated $\rho\simeq0.39$ by relating the average strain interval between plastic drops in steady state to the distribution of forward barriers~\cite{SM}. These values are therefore mutually consistent.

Once $\rho\,a^2\Delta\epsilon_0$ is thus determined, Eq.~(\ref{eq:model}) provides parameter-free predictions for both Bauschinger tests, i.e., forward and backward loading from zero-stress states. The resulting curves (Fig.~\ref{fig:bauschinger}, solid green and red lines), do match strikingly well the corresponding stress-strain relations, up to at least 5\% of strain. It establishes that the Bauschinger effect does result from the forward-backward asymmetry, in unloaded states, between the small barrier tails, i.e., for $\Delta\tau_{xy}^c\lesssim0.5$, which corresponds to the macroscopic stress change over the fitted strain range.

The remarkable ability of our model to jointly account for these three mechanical tests supports that its core assumption (barriers are mechanically shifted by macroscopic stress up to instabilities) is quite reasonable up to strains about a few percents. This legitimates using the model itself to understand how full unloading (down to zero stress) leads to the small barrier distribution asymmetry which we have just shown to be responsible for the Bauschinger effect. We thus plot in Figs.~\ref{fig:barriers}(e) and ~\ref{fig:barriers}(f) (black dashed lines) the barrier distribution $P^{\rm (m)}$ the model predicts after full unloading: it is merely the steady state distribution, translated by $\overline{\tau}_{xy}^{\rm flow}$ along the $x$ axis, and truncated to reflect the elimination of instable barriers. $P^{\rm (m)}$ is multiplied by an arbitrary factor to better show how it departs from the measured distribution, $P^{\rm (u)}$.
Since, for all $|\Delta\tau_{xy}^c|\gtrsim1$, $P^{\rm(m)}$ falls right atop $P^{\rm (u)}$, we conclude that the assumed elastic shift of barriers is a very reasonable assumption away from threshold.

The model remarkably predicts an inversion of small barrier tails during unloading, as observed, yet with two discrepancies. It overestimates the growth of the backward barrier density near threshold, expectedly due to the neglect of mechanical noise, which facilitates the crossing of small barriers, hence requires $P^{\rm (u)}$ to essentially vanish at $\Delta\tau_{xy}^c=0$~\cite{LemaitreCaroli2007,LemaitreCaroli2009}. As for the forward barrier density, the model predicts the appearance, during unloading, of a gap over $\Delta\tau_{xy}^c<\overline{\tau}_{xy}^{\rm flow}$. In this range, remarkably, the measured $P^{\rm (u)}$ does present a pseudogap, i.e., a weak initial growth compared with its rise beyond $\overline{\tau}_{xy}^{\rm flow}$. But it does not strictly vanish, presumably due to rejuvenation and/or noise associated with the small plastic activity during unloading. This analysis suggests that the elastic shift of barriers up to instabilities is the main drive behind the inversion in small barrier densities during unloading, while the neglected effects, rejuvenation and noise, which arise from unloading-induced plasticity, are only mitigating factors.

To test this interpretation, we compute the model predictions for Bauschinger's tests, when replacing the initial (unloaded) density $P^{\rm (u)}$ by $P^{\rm (m)}$. The predicted response curves are displayed in Fig.~\ref{fig:bauschinger} (dashed lines): reloading is strictly elastic; backward loading is the continuation of unloading. These curves clearly exhibit a Bauschinger effect of very reasonable amplitude, although slightly overestimated. This unambiguously confirms that the discrepancies previously identified betwen $P^{\rm (m)}$ and $P^{\rm (u)}$ only reflect compensation mechanisms, while the elastic shift (up to instabilities) hypothesis at the basis of the model captures the core mechanism responsible for unloading-induced inversion of the small barrier asymmetry leading to the Bauschinger effect.

Additionally, according to the model: (a) for any finite unloading, a gap opens in the forward barrier distribution, hence reloading is pure elastic; (b) from its very onset, unloading initiates reverse plasticity and softening. The model therefore predicts that the Bauschinger effect exists at partial unloading levels, and that the associated contrast grows with the decreasing stress. This is unambiguously confirmed by simulations~\cite{SM}. While the (pseudo)-gap formation [(a)] results merely from the stability condition $\Delta\tau_{xy}^c\ge0$, reverse softening originates from the remarkable property that the steady flow \emph{reverse} barrier distribution vanishes only at threshold.
 This feature is expected to result from mechanical noise~\cite{LemaitreCaroli2006,LemaitreCaroli2007}, which causes local stress to diffuse over the whole stability domain~\cite{LemaitreCaroli2009}. The Bauschinger effect thus appears to be an indirect consequence of mechanical noise.

This work has shown that strain induces a history-dependent polarization of local yield thresholds in an amorphous solid under AQS shear. The Bauschinger effect then appears to arise because the backward-yielding barrier distribution vanishes only (and quasilinearly) at threshold, so that unloading causes reverse plasticity of growing amplitude (i.e. softening), jointly with the emergence of a pseudogap in the forward barrier distribution, guaranteeing nearly elastic reloading. Although we used a 2D model, we expect these conclusions to carry over to 3D since our main qualitative results (forward-reverse symmetry breaking, and presence of near-threshold reverse barriers) are not dimension dependent.

\begin{acknowledgments}
M.L. and S.P. acknowledge the support of French National Research Agency
through the JCJC project PAMPAS under Grant No. ANR-17-CE30-0019-01.
\end{acknowledgments}


%

\bigskip
\bigskip
\bigskip
\bigskip
\bigskip
\bigskip
\bigskip
\bigskip

\onecolumngrid
\begin{center}
\textbf{\large Supplemental Material: ``Origin of the Bauschinger effect in amorphous solids''}
\end{center}
\twocolumngrid

\setcounter{equation}{0}
\setcounter{figure}{0}
\setcounter{page}{1}

\renewcommand{\bibnumfmt}[1]{[SM#1]}
\renewcommand{\theequation}{SM\arabic{equation}}
\renewcommand{\thefigure}{SM\arabic{figure}}

\section*{Accessing a preparation-independent steady state }

\begin{figure}[b]
\includegraphics[width=0.95\columnwidth]{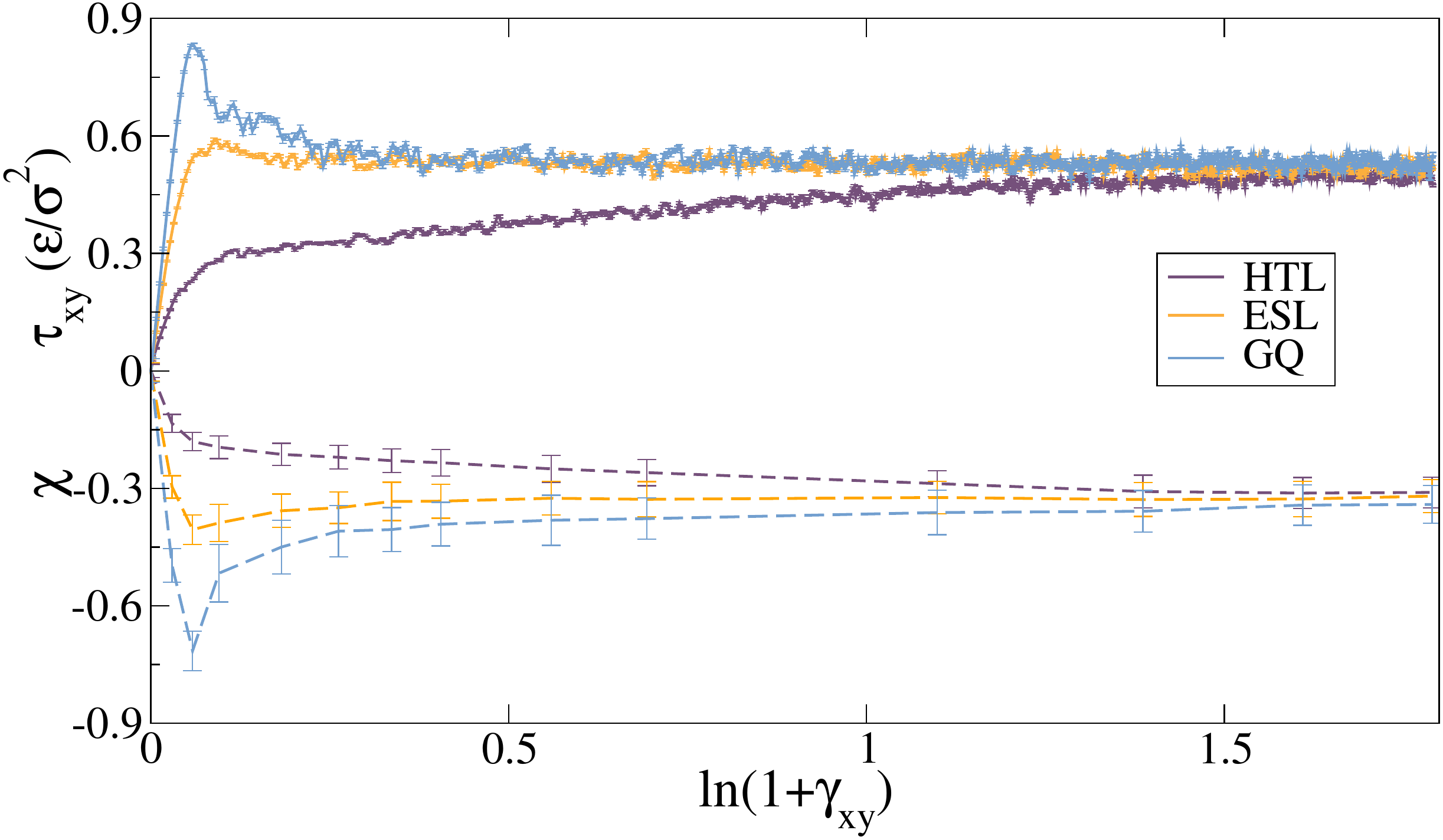}
  \caption{\label{fig:convergence}
 For our three different initial ensembles: mean stress $\tau_{xy}$ and mean barrier polarizations $\chi$ vs $\ln(1+\gamma_{xy})$, with $\gamma_{xy}$ the linear macroscopic strain.}
\end{figure}

We show in Fig.~\ref{fig:convergence}, the mean stress-strain $\tau_{xy}$ and mean barrier polarizations $\chi$ curves starting from three very different initial ensembles~\cite{barbot_local_2018}: the first two are obtained from instantaneous quenches from resp. a high temperature liquid (HTL, at $T=7.8T_{MCT}$ where $T_{MCT}$ is the temperature of the mode-coupling transition), and an equilibrated supercooled liquid (ESL, at $0.95T_{MCT}$); the third one is obtained by a gradual quench (GQ), at a rate $\dot T=0.32\times10^{-6}$ across the glass transition, which allows the system to equilibrate down to a relaxation timescale of order $T_g/\dot T\simeq10^{6}$, with $T_g$ the glass transition temperature. As seen for the GQ system, when starting from a tempered and hard glass, the early plastic response displays strain-softening, which is accompanied by transient localization~\cite{BLLVP_SB}. When starting from a very poorly tempered, very soft glass (our HTL ensemble), the erasure of the initial state shows up as a strain-hardening effect. The evolution of $\chi$ is roughly opposite of $\tau_{xy}$, since the mean stress tends to bring local packings closer, on average, to forward-yielding instabilities. All systems eventually reach the same steady flow ensemble, which we use as a starting point of our analysis.

\section*{Absence of symmetry in the barrier distribution when $\chi=0$}

\begin{figure}[h!]
\includegraphics[width=0.95\columnwidth]{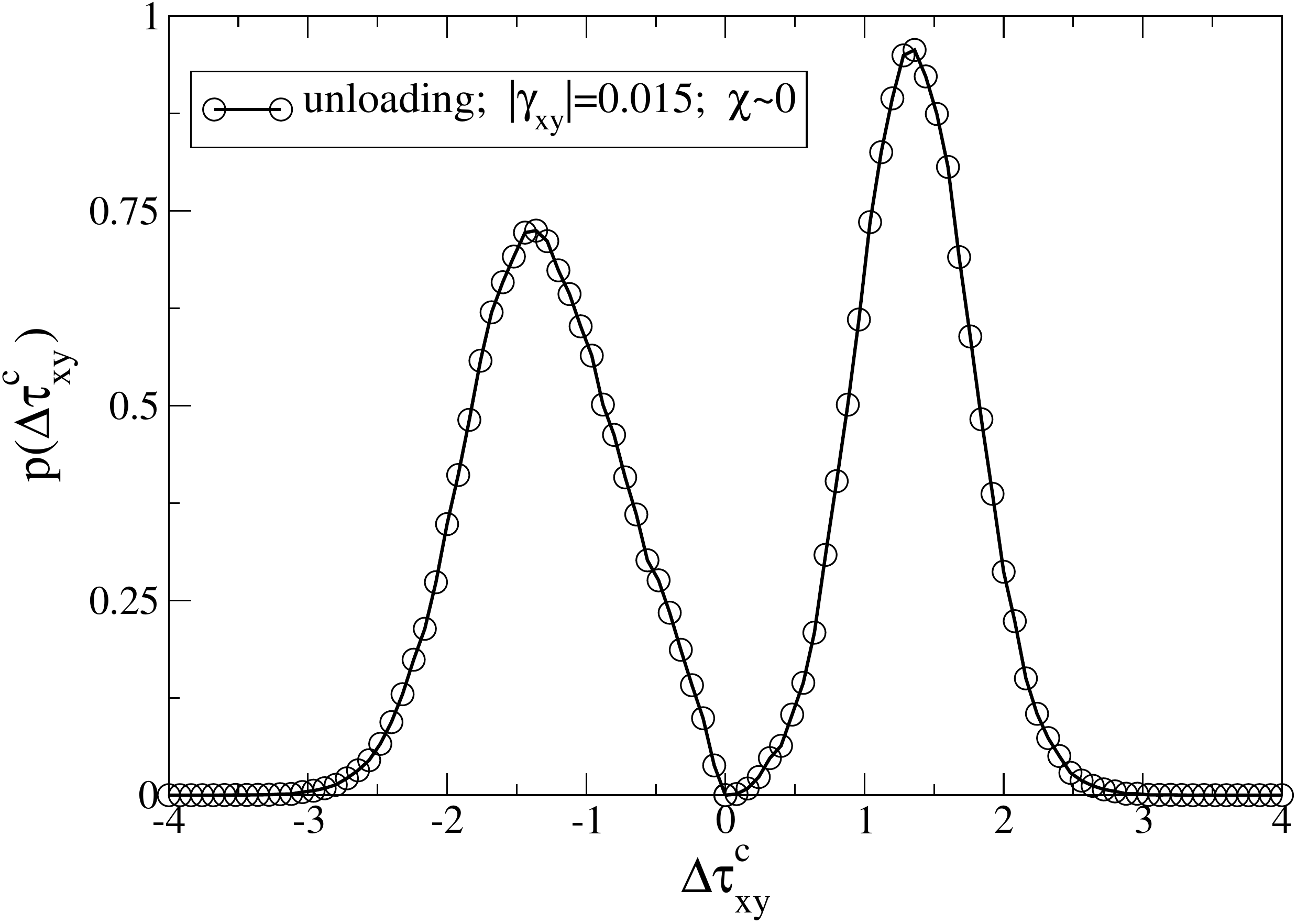}
  \caption{\label{fig:symmetry}
 Residual strength $\Delta\tau_{xy}^c$ distribution along the forward/backward loading direction in the unloading branch at $\gamma_{xy}=0.015$ that corresponds to mean barrier polarizations $\chi\approx0$.}
\end{figure}

On figure~\ref{fig:symmetry}, we report the forward and backward barrier distributions at $\gamma_{xy}=0.015$ (from the zero stress state, i.e. $\approx-3$\% of unloading), a strain at which $\chi\approx0$ (see Fig.~1 in the Article). Clearly these distributions are not symmetric, although they present almost identical $\langle\Delta\tau^c\rangle$ values.

\section*{Absence of non-trivial stress asymmetry}

\begin{figure}
\includegraphics[width=0.95\columnwidth]{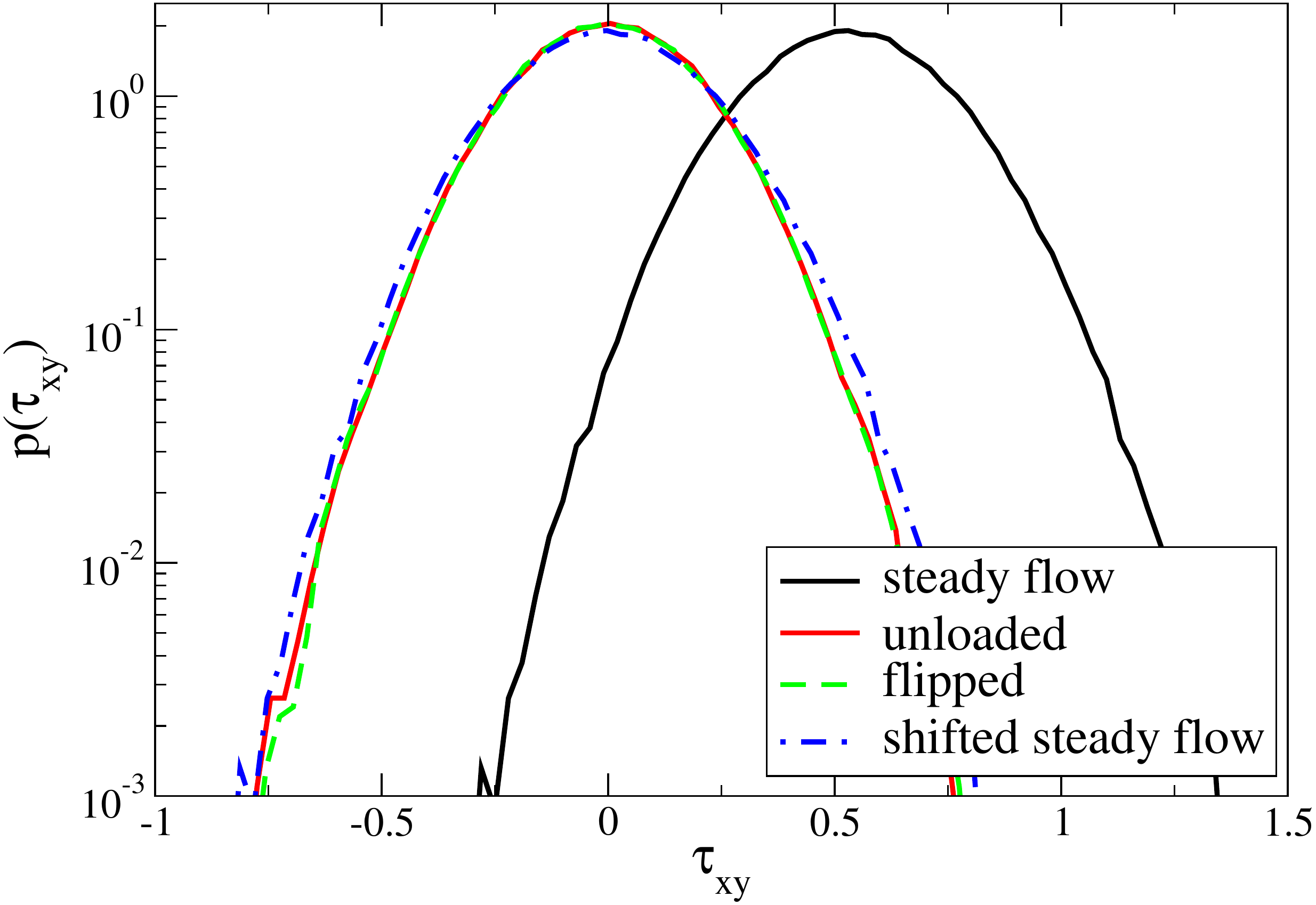}
\caption{\label{fig:stress} The distribution of inclusion stresses in flow (black) and unloaded (red) states; the latter distribution is also plotted after the $x\to-x$ transformation (green) to show that it is nearly symmetric. The steady state distribution shifted by the flow stress is also very similar (blue).}
\end{figure}

In continuum theories of plasticity, the Bauschinger effect is classically interpreted as arising from ``microstresses'', i.e. presumed local excesses of negative stress, that would cause certain regions to be closer to reverse yielding~\cite{rice_structure_1970,Asaro-ActaMetal75,mughrabi_dislocation_1988}. Here, we would like to emphasize that this interpretation does not apply to our systems, since the distribution of stress in the tested inclusions, which is reported in Fig.~\ref{fig:stress}, is nearly perfectly symmetric in unloaded states, and corresponds to the elastic shift by the mean stress $\overline\tau_{xy}$ of its counterpart in flow states.

\section*{Event samples and estimation of the typical strain release}

\begin{figure}
\includegraphics[height=0.95\columnwidth]{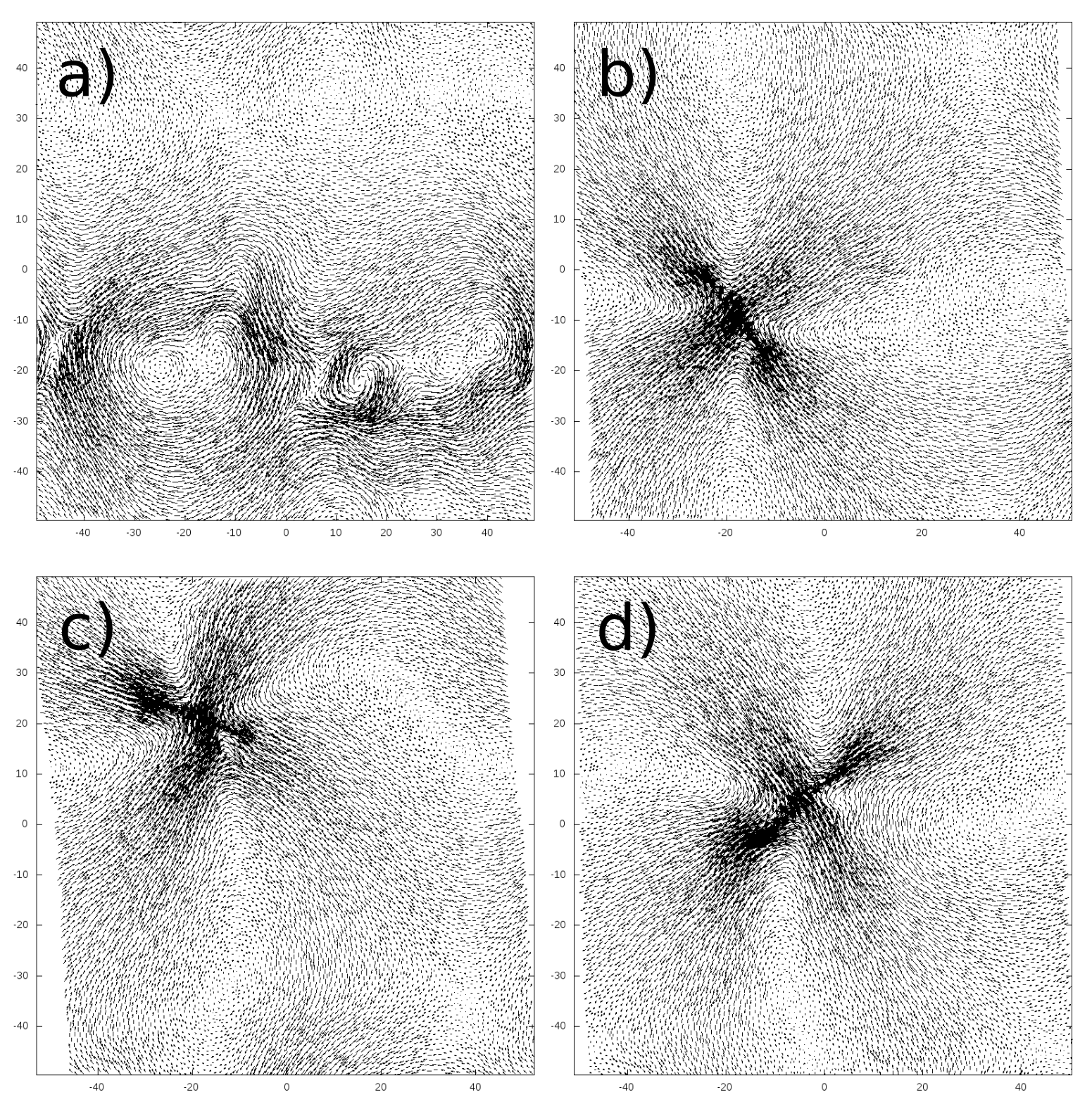}
\caption{\label{fig:dispfields} Example of displacement fields for plastic drops under different types of loading conditions. (a) in steady state, an event is typically an avalanche; in contrast, during unloading (b), reverse loading (c), or re-loading (d), typical events are well separated Eshelby-like single zone flips.}
\end{figure}

In Fig.~\ref{fig:dispfields}, we display typical displacement fields obtained from the first plastic events observed under different conditions: (a) steady flow; (b) unloading from steady state; (c) backward loading from zero stress (which is just the continuation of unloading) and (d) re-loading after full (zero stress) unloading. In steady flow, we clearly see that a typical plastic drop corresponds to a system-spanning avalanche; in the three other tests we usually observe independent Eshelby-like events. Note, however, that reverse loading, which is the continuation of unloading, occasionnally features plastic events that may combine a few spatially separated rearrangements due to the reverse polarization (higher amplitude) of the low-$\Delta\tau^{c}$ density in the backward direction.

We estimate the typical strain release $\Delta\epsilon_0$ by computing the corresponding local strain from a coarse-graining procedure describe in \cite{barbot_local_2018,BLLVP_SB} over a the typical test domain size $a=5$. This analysis is performed only for isolated plastic events and gives values lying in the 0.016--0.028 range, which corresponds to $a^2\Delta\epsilon_0$ in the 0.4--0.7 range.

\section*{Estimating $\rho$ from the average length of elastic branches in steady state}

Let us consider a configuration taken from steady state, at the end of a plastic stress drop. Previous studies~\cite{MaloneyLemaitre-PRL04b,karmakar_predicting_2010} have shown that, at such a point, the length of the following elastic branch is already determined by the local packing which first reaches instability when convected by external loading. Consistently with our simple plasticity model, we assume that: (i) in a system of volume $V=L^2$, the instability arises among one of the $N=\rho\,L^2$ barriers that are independent, distributed according to the steady state forward distribution $p$ reported on Fig.~2-(e) and (f) in the article\\
(ii) at each point, the local distance to threshold shifts as $\Delta\tau_{xy}^c(\delta\gamma_{xy})=\Delta\tau_{xy}^c(0)-\mu\delta\gamma_{xy}$ up to the first instability.\\
Under these assumptions, $\delta\gamma_{xy}>\gamma^{*}$ iff all $N$ barriers verify $\Delta\tau_{xy}^c(0)>\mu \gamma^{*}$, which occurs with probability:
\begin{equation}
\mathcal{P}(\delta\gamma>\gamma^{*})=\left(\int_{\mu\,\gamma^{*}}^{\infty}\d\delta\tau\,p(\delta\tau)\right)^N
\end{equation}

Since the density of $\delta\gamma$ is the derivative $-\mathcal{P}'(\delta\gamma)$, the average strain interval is:
\begin{equation}
  \begin{split}
    \langle\delta\gamma\rangle&=-\int_0^\infty\d x\,x \mathcal{P}'(x)\\
    &=\int_0^\infty\d x\,\mathcal{P}(x)
  \end{split}
\end{equation}
We measure $\langle\delta\gamma\rangle$ and $p$ independently and then find the value of $N$ which fits the above two relations. This yields $\rho\simeq0.39$.

\section*{Bauschinger effect at partial unloading levels}

\begin{figure}
   \includegraphics[width=0.95\columnwidth]{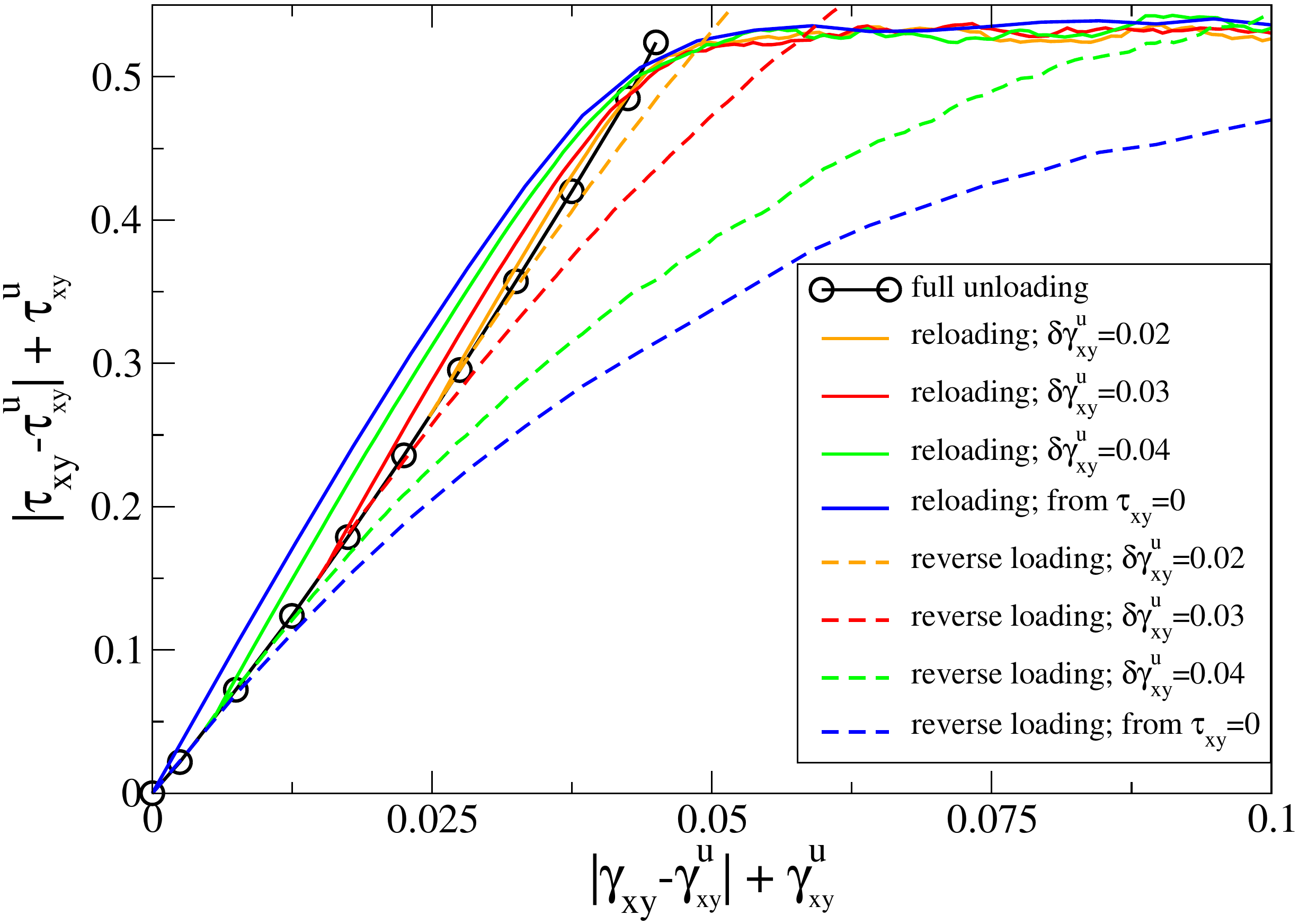}
\caption{\label{fig:bauschinger_partial} Mean stress vs strain for different responses: full unloading from steady flow (black symbols); after full or partial unloading (by $\delta\gamma^{u}_{xy}$) from steady state, during re-loading (continuous lines), and backward loading (after inversion about $(\gamma_{xy}^{u},\tau_{xy}^u)$, dashed lines). In all cases, strain is measured with reference to the zero-stress state.}
\end{figure}

On figure~\ref{fig:bauschinger_partial}, we report the forward and backward mechanical response, starting at a few levels of unloading from steady flow, ranging from $\delta\gamma_{xy}^{u}=-0.02$ (down from steady state) to fully unloaded (zero stress, $\delta\gamma_{xy}^{\rm full}\approx-0.045$) configurations. In unloaded states, the strain measured with respect to the zero stress state is $\gamma_{xy}^{u}=\delta\gamma_{xy}^{\rm full}-\delta\gamma_{xy}^{u}$. To evidence the response contrast, the re-loading data is plotted after the inversion about the point $(\gamma_{xy}^{u},\tau_{xy}^u)$, which amounts to plotting in all cases $|\tau_{xy}-\tau_{xy}^u|+\tau_{xy}^u$ vs $|\gamma_{xy}-\gamma_{xy}^u|+\gamma_{xy}^u$. These curves unambiguously confirm our prediction that the Bauschinger effect exists at finite unloading, and that the associated contrast between forward and backward responses grows with the increasing unloading level.

\end{document}